\documentclass{elsart}

\usepackage{amssymb}
\usepackage{epsfig}

\def\H{H\hskip-8.5pt/\hskip2pt}

\def\coeff#1#2{{\textstyle{#1\over #2}}}

\def\VEV#1{\left\langle #1\right\rangle}

\def\lsim{\mathrel{\mathpalette\@versim<}}

\def\gsim{\mathrel{\mathpalette\@versim>}}

\def\Tr{{\rm Tr}\,}

\begin{document}

\begin{frontmatter}



\title{Theoretical and Phenomenological Aspects 
of CPT Violation}


\author{Nick E. Mavromatos}

\address{King's College London, Department of Physics, Theoretical 
Physics, Strand, London WC2R 2LS, U.K.}

\begin{abstract}
I review briefly various models and ways of 
Quantum-Gravity induced CPT violation, 
and discuss in some detail their phenomenology, in particular 
precision CPT tests 
in neutral mesons, and hydrogen/antihydrogen spectroscopy. 
As I shall argue, severe
constraints can be placed in CPT violating parameters,
with sensitivities that can safely exclude models  
with effects suppressed by a single power of Planck mass.

\end{abstract}

\end{frontmatter}

\section{Introduction: CPT Theorem and its Violation}
\label{}

Any complete theory of quantum gravity is bound to address
fundamental issues, directly related to the 
emergence of space-time and its structure at energies 
beyond the Planck energy scale $M_P \sim 10^{19} $ GeV.  
From our relatively low energy experience so far, we are lead to expect
that a theory of quantum gravity should respect most of the fundamental 
symmetries of particle physics, 
that govern the standard model of electroweak and strong 
interactions: Lorentz symmetry and CPT invariance,
that is invariance under the combined action of Charge Conjugation (C),
Parity (reflection P) and Time Reversal Symmetry (T).
Actually the latter invariance is a theorem of any local quantum field theory
that we can use to describe the standard phenomenology of particle
physics to date. The {\bf CPT theorem} can be stated as follows~\cite{cpt}: 
{\bf Any quantum theory, formulated on {\it flat space time} is symmetric
under the combined action of CPT transformations, provided the theory  
respects} (i) {\it Locality}, (ii) {\it Unitarity} (i.e. conservation of 
probability) and (iii) 
{\it Lorentz invariance}. 

If such a theorem exists, then why do we have to bother to test
CPT invariance, given that all our phenomenology up to now has been 
based on such quantum theories ? The answer to this question is 
intimately linked with our understanding of {\it quantum gravity}.
First of all, the theorem is not valid (at least in its strong form) 
in highly curved ({\it singular}) 
space times,
such as black holes, or in general in space-time 
backgrounds of some quantum gravity theories 
involving the so-called {\it quantum 
space-time foam} backgrounds~\cite{wheeler}, 
that is {\it singular} quantum fluctuations of space time geometry,
such as black holes {\it etc}, with event horizons 
of microscopic Planckian size ($10^{-35}$ meters). 
Such backgrounds result in {\it apparent} violations of {\it unitarity}
in the following sense: there is part of 
information (quantum numbers of incoming matter) 
``disappearing'' inside the  microscopic event 
horizons, so that an observer at asymptotic infinity will have to 
trace over such ``trapped'' degrees of freedom. Thus, one faces a 
situation in which an initially pure state evolves in time
to get mixed: the asymptotic states are described by density matrices,
defined as follows: 
$$ \rho _{\rm out} = {\rm Tr}_{M} |\psi ><\psi|~,$$ 
where the 
trace is over trapped (unobserved) quantum states, that disappeared 
inside the microscopic event horizons in the foam. 
Such a non-unitary evolution results in the impossibility of defining 
a standard quantum-mechanical scattering matrix, connecting asymptotic
states in a scattering process:
$|{\rm out}> = S~|{\rm in}>,~S=e^{iH(t_f - t_i)} $,
where $t_f - t_i$ is the duration of the scattering (assumed much longer than
other time scales in the problem). Instead, in 
foamy situations, one can define 
an operator that connects asymptotic density matrices~\cite{hawking}:
$$\rho_{\rm out} \equiv {\rm Tr}_{M}| {\rm out} ><{\rm out} | 
= \$ ~\rho_{\rm in},~ \qquad \$ \ne S~S^\dagger $$ where 
the lack of factorization is attributed to the apparent loss of unitarity of the effective low-energy theory, defined as the part of the theory
accessible to low-energy observers who perform scattering experiments.
This defines what we mean by {\it particle phenomenology} in such  situations.

The \$ matrix is {\it not invertible}, and this reflects the 
effective unitarity loss. It is this property, actually, that leads to a 
{\bf violation of CPT invariance} (at least in its strong form) in such a 
situation~\cite{wald}, since one of the requirements of 
CPT theorem (unitarity) is violated: 
{\bf In an open (effective) quantum theory, 
interacting with an environment, e.g. quantum gravitational,  where} 
$ \$ \ne SS^\dagger $, {\bf CPT invariance is violated, 
at least in its strong form}.
The proof is based on elementary quantum mechanical concepts and the above-mentioned
non-invertibility of \$, but will be omitted here due to lack of 
space~\cite{wald}. Another reason for CPT violation (CPTV) 
in quantum gravity is {\it spontaneous breaking of Lorentz symmetry}, 
without necessarily 
implying decoherence. This may also occur in string theory and other
models. In certain circumstances one may also violate locality, 
e.g. of the type advocated in \cite{lykken} to explain 
observed neutrino 
physics `anomalies', 
but we shall not discuss this case here.

The CPT violating effects can be estimated naively to be 
strongly suppressed, and thus inaccessible - for all practical purposes
-to current, or immediate future, low-energy experiments.
Indeed, naively, Quantum Gravity (QG) has a dimensionful constant:
$G_N \sim 1/M_P^2$, where $M_P =10^{19}$ GeV is the Planck scale. 
Hence, CPT violating 
and decoherening 
effects may be expected to be suppressed
by $E^3/M_P^2 $, where $E$ 
is a typical energy scale of the low-energy 
probe. However, there may be cases where loop resummation and other 
effects 
in theoretical 
models 
may result in much larger CPT-violating effects of order: 
$\frac{E^2}{M_P}$.
This happens, for instance, in some loop gravity approaches to 
QG, or some non-equilibrium stringy models of space-time
foam involving 
open string excitations. Such large effects 
can lie within the sensitivities of 
current or immediate future experimental facilities
(terrestrial and astrophysical).
Below we shall describe a few such sensitive probes, starting from 
neutral kaon decays.

\section{Quantum Gravity Decoherence and CPT Violation in Neutral Kaons}

QG may induce decoherence and oscillations 
$K^0 \to {\overline K}^0$~\cite{ehns,lopez}.  
The modified evolution equation for the respective density matrices
of neutral kaon matter can be parametrized as follows~\cite{ehns}: 
$$\partial_t \rho = i[\rho, H] + \delta\H \rho~,$$ 
where 
$$H_{\alpha\beta}=\left( \begin{array}{cccc}  - \Gamma & -\coeff{1}{2}\delta \Gamma
& -{\rm Im} \Gamma _{12} & -{\rm Re}\Gamma _{12} \\
 - \coeff{1}{2}\delta \Gamma
  & -\Gamma & - 2{\rm Re}M_{12}&  -2{\rm Im} M_{12} \\
 - {\rm Im} \Gamma_{12} &  2{\rm Re}M_{12} & -\Gamma & -\delta M    \\
 -{\rm Re}\Gamma _{12} & -2{\rm Im} M_{12} & \delta M   & -\Gamma
\end{array}\right) $$ and 
$$ {\delta\H}_{\alpha\beta} =\left( \begin{array}{cccc}
 0  &  0 & 0 & 0 \\
 0  &  0 & 0 & 0 \\
 0  &  0 & -2\alpha  & -2\beta \\
 0  &  0 & -2\beta & -2\gamma \end{array}\right)~.$$
Positivity of $\rho$ requires:
$\alpha, \gamma  > 0,\quad \alpha\gamma>\beta^2$.
Notice that 
$\alpha,\beta,\gamma$ violate CPT, as they do not commute
with a CPT operator $\Theta$~\cite{lopez}:  
$\Theta = \sigma_3 \cos\theta + \sigma_2 \sin\theta$,$~~~~~[\delta\H_{\alpha\beta}, \Theta ] \ne 0$.

An important remark is now in order. 
We should distinguish two types of CPTV:
(i) CPTV within Quantum Mechanics~\cite{takeuchi}:
$\delta M= m_{K^0} - m_{{\overline K}^0}$, $\delta \Gamma = \Gamma_{K^0}-
\Gamma_{{\overline K}^0} $. 
This could be  due to (spontaneous) Lorentz violation (c.f. below).\\
(ii) CPTV through decoherence $\alpha,\beta,\gamma$ 
(entanglement with QG `environment', leading to modified 
evolution for $\rho$ and  $\$ \ne S~S^\dagger $). 

\begin{table}[thb]
\begin{center}
\begin{tabular}{lcc}
\underline{Process}&QMV&QM\\
$A_{2\pi}$&$\not=$&$\not=$\\
$A_{3\pi}$&$\not=$&$\not=$\\
$A_{\rm T}$&$\not=$&$=$\\
$A_{\rm CPT}$&$=$&$\not=$\\
$A_{\Delta m}$&$\not=$&$=$\\
$\zeta$&$\not=$&$=$
\end{tabular}
\caption{Qualitative comparison of predictions for various observables
in CPT-violating theories beyond (QMV) and within (QM) quantum mechanics.
Predictions either differ ($\not=$) or agree ($=$) with the results obtained
in conventional quantum-mechanical CP violation. Note that these frameworks can
be qualitatively distinguished via their predictions for $A_{\rm T}$, $A_{\rm
CPT}$, $A_{\Delta m}$, and $\zeta$.}
\label{Table2}
\end{center}
\hrule
\end{table}

The important point is that the two types of CPTV can be {\bf disentangled 
experimentally}~\cite{lopez}. 
The relevant observables are defined as $ \VEV{O_i}= {\rm Tr}\,[O_i\rho] $.
For neutral kaons, one looks at decay asymmetries 
for $K^0, {\overline K}^0$, defined as: 
$$A (t) = \frac{
    R({\bar K}^0_{t=0} \rightarrow
{\bar f} ) -
    R(K^0_{t=0} \rightarrow
f ) }
{ R({\bar K}^0_{t=0} \rightarrow
{\bar f} ) +
    R(K^0_{t=0} \rightarrow
f ) }~,$$ 
where $R(K^0\rightarrow f) \equiv \Tr[O_{f}\rho (t)]=$ denotes the decay rate
into the final state $f$ (starting from a pure $ K^0$ state at $t=0$).

In the case of neutral kaons, one may consider the 
following set of asymmetries:
(i) {\it identical final states}: 
$f={\bar f} = 2\pi $: $A_{2\pi}~,~A_{3\pi}$,
(ii) {\it semileptonic} : $A_T$
(final states $f=\pi^+l^-\bar\nu\ \not=\ \bar f=\pi^-l^+\nu$), $A_{CPT}$ (${\overline f}=\pi^+l^-\bar\nu ,~ f=\pi^-l^+\nu$), 
$A_{\Delta m}$.  Typically, for instance when final states are $2\pi$,
one has  a time evolution of the decay rate $R_{2\pi}$: 
$ R_{2\pi}(t)=c_S\, e^{-\Gamma_S t}+c_L\, e^{-\Gamma_L t}
+ 2c_I\, e^{-\Gamma t}\cos(\Delta mt-\phi)$, where 
$S$=short-lived, $L$=long-lived, $I$=interference term, 
$\Delta m = m_L - m_S$, $\Gamma =\frac{1}{2}(\Gamma_S + \Gamma_L)$. 
One may define the {\bf Decoherence Parameter}
$\zeta=1-{c_I\over\sqrt{c_Sc_L}}$, as a measure 
of quantum decoherence induced in the system. 
For larger sensitivities one can look 
at this parameter in the presence of a 
regenerator~\cite{lopez}. 
In our decoherence scenario, $\zeta$ depends primarily on $\beta$,
hence the best bounds on $\beta$ can be placed by 
implementing a regenerator~\cite{lopez}.

\begin{figure} 
\centering
  \epsfig{file=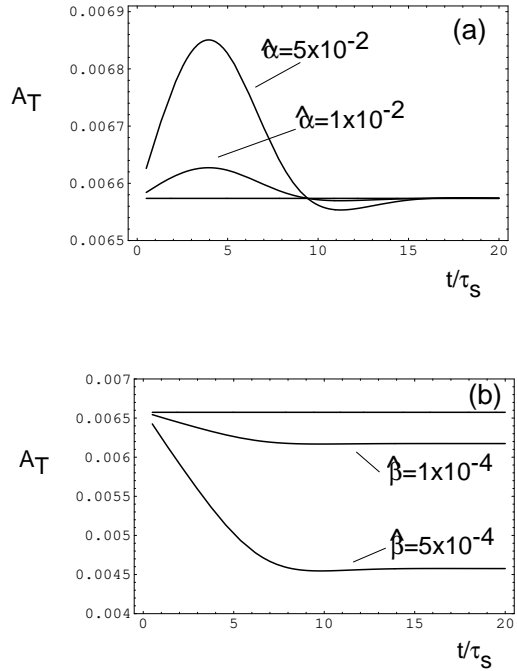, width=0.5\textwidth}
\caption{A typical neutral kaon decay asymmetry $A_T$~\cite{lopez} 
indicating the 
effects of quantum-gravity induced decoherence.} 
\label{AT}
\end{figure}

The experimental tests (decay asymmetries) 
that can be performed in order to disentangle
decoherence from quantum mechanical CPT violating effects 
are summarized in table \ref{Table2}. 
In figure \ref{AT} we give a typical profile of a decay asymmetry,
that of $A_{T}$~\cite{lopez},
from where bounds on QG decoherening parameters can be extracted. 
Experimentally, the best available bounds come from 
CPLEAR measurements~\cite{cplear} 
$\alpha < 4.0 \times 10^{-17} ~{\rm GeV}~, ~|\beta | < 2.3. \times
10^{-19} ~{\rm GeV}~, ~\gamma < 3.7 \times 10^{-21} ~{\rm GeV} $,
which are not much different from 
theoretically expected values 
$\alpha~,\beta~,\gamma = O(\xi \frac{E^2}{M_{P}})$.

\section{Spontaneous Violation of Lorentz Symmetry
and (Anti)Hydrogen}

A second possibility for CPTV effects arises if the Lorentz symmetry 
is violated {\it spontaneously}, 
but no quantum decoherence or unitarity loss necessarily occurs. 
Such a situation may be envisaged in some 
string theory (non supersymmetric) models, where some 
tensorial fields acquire vevs 
$<T_{\mu_1 \dots \mu_n}>  \ne 0~.$ 
This will result in a
spontaneous breaking of Lorentz symmetry by 
(exotic) string vacua, implying a
{\bf modified Dirac equation (MDE)} for fermions
in the so-called Standard Model Extension (SME)~\cite{sme,kostelecky}.
In view of the recent `massive' production of antihydrogen 
(${\overline H}$ ) at 
CERN \cite{cernhbar}, 
which implies that 
interesting direct tests of CPT invariance using ${\overline H}$
are to be expected in the near future, we   
consider for our purposes here the specific case of MDE 
for Hydrogen $H$ (anti-hydrogen ${\overline H}$). 
Let
the spinor $\psi$  represent the electron  
(positron) with charge $q=-|e| (q=|e|)$ 
around a proton (antiproton) 
of charge $-q$. Then the MDE reads:
$$\left( i\gamma^\mu D^\mu - M -  
a_\mu \gamma^\mu - b_\mu \gamma_5 \gamma^\mu -
\frac{1}{2}H_{\mu\nu}\sigma ^{\mu\nu}
+  ic_{\mu\nu}\gamma^\mu D^\nu + id_{\mu\nu}\gamma_5\gamma^\mu D^\nu  
\right)\psi =0,$$
where $D_\mu = \partial_\mu - q A_\mu$, $A_\mu = (-q/4\pi r, 0)$ Coulomb 
potential. The parameters $a_\mu~, b_\mu $ induce 
CPT and Lorentz violation, while the parameters $c_{\mu\nu}, d_{\mu\nu},  
H_{\mu\nu} $ induce Lorentz violation only.

In SME models there are energy shifts between states 
$|J,I;m_J,m_I>$, with $J (I)$ denoting electronic (nuclear) 
angular momenta. Using 
perturbation theory, one finds~\cite{kostelecky}:
\begin{eqnarray} 
&&\Delta E^H (m_J, m_I) \simeq  a_0^e + a_0^p - c_{00}^e m_e - c_{00}^p m_p 
+ (-b_3^e + d_{30}^em_e + H_{12}^e)\frac{m_J}{|m_J|} + \nonumber \\
&&(-b_3^p + d_{30}^p m_p + H_{12}^p)\frac{m_I}{|m_I|}~, \nonumber 
\end{eqnarray} 
where $e$ electron; $p$ proton. The 
corresponding results for antihydrogen (${\overline H}$) 
are obtained upon:
$$a_\mu^{e,p} \rightarrow -a_\mu^{e,p}~,~ 
b_\mu^{e,p} \rightarrow -b_\mu^{e,p}~,~ 
d_{\mu\nu}^{e,p} \rightarrow d_{\mu\nu}^{e,p}~, 
H_{\mu\nu}^{e,p} \rightarrow H_{\mu\nu}^{e,p}~.$$

One may study the spectroscopy of {\bf forbidden 
transitions 1S-2S}: If CPT and Lorentz violating parameters are
constant they drop out to leading order energy shifts in free H 
(${\overline H})$.  Subleading effects are then suppressed by 
the square of the fine structure constant:
$\alpha^2 \sim 5 \times 10^{-5}$, specifically:  
$\delta \nu^H_{1S-2S} \simeq -\frac{\alpha^2b_3^e}{8\pi}$.
This is too small to be seen. 

\begin{figure}
\centering
\epsfig{file=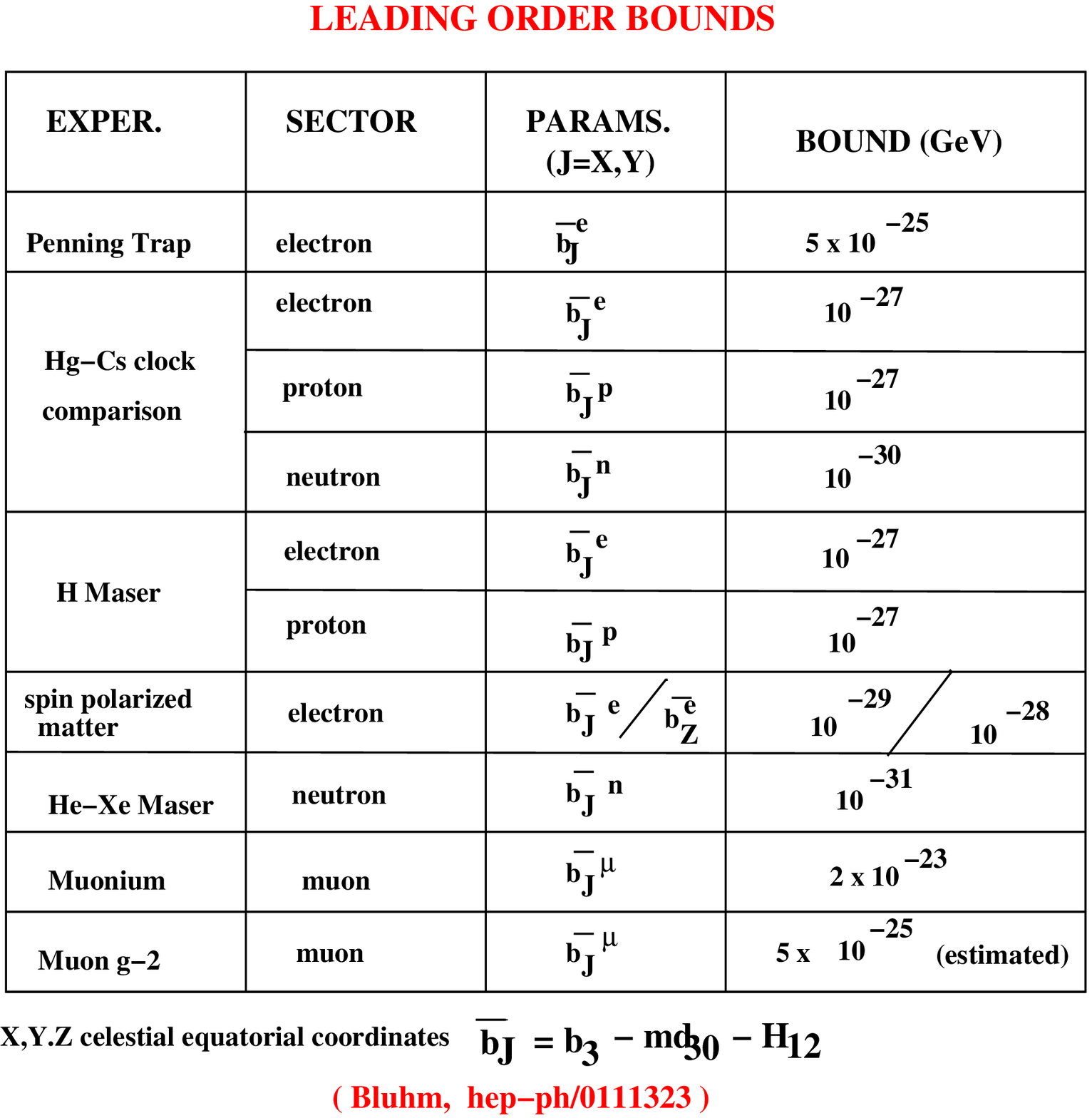, width=0.9\textwidth}
\caption{Table summarising recent bounds of CPT violating 
parameter $b$ in the Standard Model extension 
from atomic and nuclear physics spectroscopic tests 
(from Bluhm hep-ph/0111323).}
\label{bounds}
\end{figure}

But what about 
the case where atoms of $H$ (or ${\overline H}$) are 
in magnetic traps? Magnetic fields 
induce hyperfine Zeeman splittings in 1S, 2S states.
There are four spin states, mixed under the 
the magnetic field B ($|m_J,m_I>$ basis):
$|d>_n =|\frac{1}{2},\frac{1}{2}>$,
$|c>_n ={\rm sin}\theta_n|-\frac{1}{2},\frac{1}{2}> +
{\rm cos}\theta_n |\frac{1}{2},-\frac{1}{2}>$,
$|b>_n =|-\frac{1}{2},-\frac{1}{2}>$,
$|a>_n ={\rm cos}\theta_n|-\frac{1}{2},\frac{1}{2}> -
{\rm sin}\theta_n |\frac{1}{2},-\frac{1}{2}>$,  
where ${\rm tan}2\theta_n= (51 {\rm mT})/{\rm n^3B}$. The 
$|c>_1 \to |c>_2$ transitions yield dominant effects for 
CPTV~\cite{kostelecky}: 
\begin{eqnarray} 
&& \delta \nu_c^H \simeq -
\frac{\kappa (b_3^e-b_3^p-d_{30}^em_e + d_{30}^pm_p-H_{12}^e + 
H_{12}^p)}{2\pi}~, \nonumber \\  
&&\delta \nu_c^{\overline H} \simeq 
-\frac{\kappa (-b_3^e+b_3^p-d_{30}^em_e - d_{30}^pm_p-H_{12}^e + H_{12}^p)}{2\pi}~, \nonumber \\
&&\Delta \nu_{1S-2S,c} \equiv \delta \nu_c^H -\delta \nu_c^{\overline H} \simeq
-\frac{\kappa(b_3^e-b_3^p)}{\pi}~, \nonumber 
\end{eqnarray}  
where $\kappa ={\rm cos}2\theta_2 - {\rm cos}2\theta_1$,  
$\kappa \simeq 0.67 $ for $B=0.011$ T. 
Notice that $\Delta \nu_{c\to d} \simeq -2b_3^p/\pi~$, and,  
{\bf if a frequency resolution of 1 mHz is attained}, one may obtain 
a bound $|b_3| \le 10^{-27} {\rm GeV}~$. 
Other low energy atomic and nuclear physics experiments may place
stringent bounds on spatial components of the  
CPTV parameters of the SME, and are summarized
in figure \ref{bounds}~\cite{bluhm}.

We next point out that, in some stringy models 
of space time foam, interaction of string matter with space-time solitonic 
defects results in  
a modified Dirac equation of SME type but only with  
(boost sensitive) temporal components of $a_0$ 
which, however, turn out to be energy dependent~\cite{gravanis}. 
For instance, 
for protons, one has $a_0 \sim 
\xi \frac{E^3}{E-m_p}\frac{1}{M_P} ~, $
where $\xi $ depends on string interaction 
coupling and is model dependent.
The model also predicts modified Dispersion relations~\cite{emn}. 
The energy dependence of $a_0$ in this case 
implies that hyperfine Zeeman splittings 
due to external magnetic field $B$ acquire shifts 
$\Delta E \sim a_0 (E)$. Hence (say 1S level): 
$$\delta \nu^H_{1S} -\mu_N B \sim \frac{\xi}{M_P}
\frac{m_p^3}{\epsilon ^2_{1S}}\mu_N B  \sim 
\xi 10^{-21} (\frac{B}{{\rm mT}})~{\rm GeV} $$
where $\epsilon _{1S}$ is the energy level, $\mu_N$ nuclear 
magneton. $H$, ${\overline H}$ spectroscopic 
measurements may be devised to constrain
the parameter $\xi$ in $a_0$. Also, one may envisage using 
relativistic beams 
of $H$, ${\overline H}$ to enhance such CPTV effects. 

A note is appropriate at this stage 
on the {\bf frame dependence} of the above results on CPTV effects. 
If Lorentz symmetry is violated (LV) then the effects should be frame
dependent. $\Delta \nu_c^H$ depends on spatial components
of LV couplings, and so it 
is subject to sidereal variations due to Earth rotation 
(clock comparison experiments using H alone). 
Usually, in such situations 
there is a preferred frame, which might be taken to be 
the cosmic microwave background frame with velocity $w \sim 10^{-3} c$.
High precision tests are then possible, if modified dispersion 
relations for matter probes exist; such tests  
proceed via quadrupole moment measurements~\cite{urrutia}, which 
exhibit sensitivities  up to  
$10^{23} {\rm GeV} > M_P=10^{19}$ GeV for minimally suppressed QG modified
dispersion relations. 
Severe constraints on such models 
come also from astrophysics~\cite{crab} (e.g. 
Crab Nebula magnetic field measurements imply
sensitivity of some quantum gravity effects 
up to scales $10^{27}$ GeV $>> M_P =10^{19}$ GeV).

\section{Conclusions} 

There are plenty of low energy nuclear and atomic physics
experiments which yield stringent bounds in models with 
Lorentz and CPT violation. Frame dependence of Lorentz
violating (LV) effects may be  
crucial in providing such stringent experimental constraints.
Indeed, experiments from nuclear physics
(via quadrupole moment measurements) can constrain
some models of QG predicting LV modified dispersion relation of
matter probes, by exploiting appropriately  the frame dependence
of such effects. It is worthy of stressing that 
such measurements exhibit sensitivity to energy scales that 
exceed the Planck scale by several orders of magnitude, thereby  
safely excluding models
with minimal (linear) Planck scale suppression.

The recently `massive' production of 
Antihydrogen \cite{cernhbar} will undoubtedly turn out to be  
very useful in 
providing physical systems appropriate 
for placing stringent bounds 
on some of these CPTV parameters 
(relevant to spontaneous 
violation of Lorentz symmetry) 
via spectroscopic measurements 
and comparison with hydrogen results, provided 
the frequency resolution improves.
A natural question arises at this point, concerning 
the possibility of constraining CPT violating QG-induced 
decoherence parameters 
using $H$, ${\overline H}$. This remains to be seen. 
In addition, such tests may be performed 
in other low-energy probes such as slow neutrons
in the gravitational field of the Earth. 
Preliminary studies 
in this system reveal a striking formal similarity with that  
of neutral kaons, and the analysis can be easily carried
through in this case. At present, however, stringent bounds 
on the decoherening parameters cannot be placed.

Certainly, more work needs to be done,
both theoretical and experimental, 
before conclusions are reached,
but we do think that the current and immediate future 
experimental situation looks very promising
in providing important information about Planck scale Physics
from low energy high precision data.



\end{document}